\documentclass[aps,prl,twocolumn,groupedaddress]{revtex4}

\usepackage{graphicx}
\begin{document}

\title{Cavity QED with optically transported atoms}


\author{J. A. Sauer}
\author{K. M. Fortier}
\author{M. S. Chang}
\author{C. D. Hamley}
\author{M. S. Chapman}
\affiliation{School of Physics, Georgia Institute of Technology,
  Atlanta, GA 30332-0430 }

\date{\today}

\begin{abstract}

Ultracold  $^{87}$Rb atoms are delivered into a high-finesse
optical micro-cavity using a translating optical lattice trap and
detected via the cavity field. The atoms are loaded into an
optical lattice from a magneto-optic trap (MOT) and transported
1.5 cm into the cavity. Our cavity satisfies the strong-coupling
requirements for a single intracavity atom, thus permitting
real-time observation of single atoms transported into the cavity.
This transport scheme enables us to vary the number of intracavity
atoms from 1 to $>$100 corresponding to a maximum atomic
cooperativity parameter of 5400, the highest value ever achieved
in an atom--cavity system.  When many atoms are loaded into the
cavity, optical bistability is directly measured in real-time
cavity transmission.

\end{abstract}


\maketitle

Many applications in quantum information science require the
coherent and reversible interaction of \emph{single photon} fields
with material qubits such as trapped atoms.  Quantum states can be
transferred between light and matter---respectively offering long
range communication and long-term storage of quantum information.
This important paradigm is the heart of cavity QED systems, which
are largely focused on creating laboratory systems capable of
reversible matter-photon dynamics at the single photon level
\cite{BermanBook}.  To achieve this, a small high-finesse build-up
cavity is used to tremendously enhance the electric field per
photon and hence the interaction strength of a single photon with
the cavity medium (\textit{e.g.} atoms). For a single atom in the
cavity, the interaction strength is given by the single photon
Rabi frequency, $2g_0$, and coherent dynamics is achieved for
$g_0^2/(\kappa\Gamma)\gg1$, where $\kappa$ is the the cavity field
decay rate and $\Gamma$ is the atomic spontaneous emission rate.

There have been spectacular recent successes in cavity QED
research brought about by the merging of optical cavity systems
with ultracold neutral atoms \cite{cqedreviewopt}, including
real-time observation \cite{Mabuchioptlet,Hood1,Rempe1} and
trapping \cite{Ye,Hood2,Rempe2,McKeever} of single atoms in
optical cavities, real-time feedback control on a single atom
\cite{Rempe3}, and single photon generation \cite{Rempe4,Rempe5}.
Together with the remarkable experimental work in microwave cavity
QED \cite{cqedreviewmic} and the future prospects for cavity QED
with trapped ions \cite{Waltherioncavity, Blattioncavity}, the
field is well-poised to contribute significantly to the
development of quantum information science. Indeed, current cavity
QED parameters are sufficient for existing quantum gate protocols
with fidelities $>99.9\%$ percent \cite{PGCZ,Walther,You1,You2},
and the systems are principally limited by the lack of a scalable
atomic trapping system to provide adequate control over atom
motional degrees of freedom.

Our strategy for overcoming this limitation is to employ optical
dipole trapping fields independent from the cavity and orthogonal
to its axis as illustrated in Fig.\ 1 \cite{apstalk}. Atoms are
trapped in the anti-nodes of the standing wave formed by two
focused counter-propagating laser beams.  Translating the standing
wave controllably introduces atoms in this 1-D chain in and out of
the cavity mode. This trapping geometry will ultimately allow
entanglement of $>$100 distiguishable atoms in two such parallel
chains or in a single chain using nearest-neighbor interactions.

\begin{figure}
\includegraphics{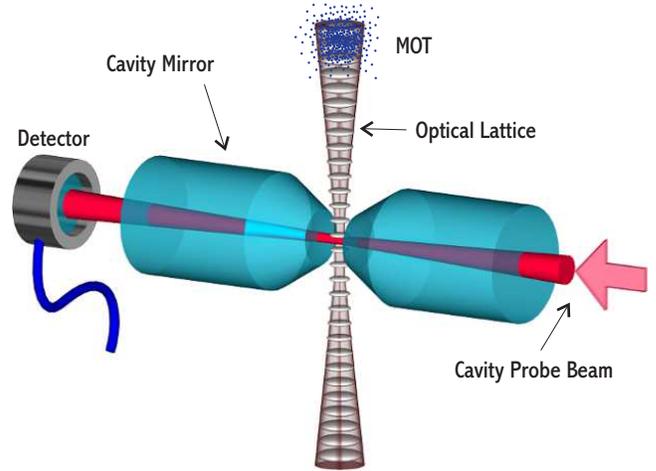}
\caption{ Two counter-propagating laser beams focused through the
cavity in the vertical direction produce an optical lattice.
Translating the lattice transports atoms collected in the MOT into
the cavity mode below.}
\label{fig:cavity_scheme}
\end{figure}

In this Letter, we describe an experimental realization of a
cavity QED system with optically trapped and transported atoms.
Single and multiple atoms are transported into the cavity and
detected---in the latter case, we observe optical bistability in
the cavity output due to the cooperative interaction of the atoms
with the cavity field. By adjusting the initial loading conditions
of the optical traps, the number of intracavity atoms, $N$, can be
varied from 1-106 atoms. Our unique combination of strong coupling
parameters and large maximum intracavity atomic density ($10^{10}$
cm$^{-3}$), provide for a measured atomic cooperativity,
$C$=$Ng_0^2/(\kappa\Gamma)$, up to 5400, the highest value ever
achieved in an atom-cavity system. The cooperativity parameter is
an important figure of merit both for single atom ($N=1$)
protocols and for many-atom, ensemble based protocols including
light storage and single photon generation
\cite{Zoller,SoMo,Kuga,Lukin,kuzmich}.

The experiment begins 1.5 cm above the optical cavity with a
laser-cooled sample of $^{87}$Rb atoms collected in a beam-loaded
MOT as previously described in \cite{storagering}.  The cavity
itself consists of two 1 mm diameter super-polished mirrors with
10 cm radii of curvature separated by 75 $\mu$m.  The measured
finesse of the cavity is $\textsl{F}=420,000$. The relevant cavity
QED parameters for our system are $(g_0,\kappa,\Gamma)/(2\pi) =
(27,2.4,6)$ MHz, placing our system well into the strongly coupled
regime and yielding a single atom cooperativity parameter,
$C_1$=$g_0^2/(\kappa\Gamma)=51$.

Maintaining the cavity on resonance requires cavity length
stability better than $\Delta L<0.1\lambda/\textsl{F}\sim 200$ fm
($\lambda=780$ nm is the resonant wavelength of the cavity
) and
represents a significant challenge.   To achieve the required
stability while providing the necessary length adjustability, both
mirrors are mounted directly on a flat rectangular piezoelectric
transducer (PZT) which in turn is attached to a copper block. The
block is suspended inside a vacuum chamber by 4 Cu-Be springs to
provide an \emph{in-vacu} isolation system with a 3.5 Hz
resonance. Both the copper block and the PZT have holes drilled
vertically to accommodate the optical lattice beams. The length of
the cavity is controlled by applying a voltage to the PZT.  The
isolation system together with low-noise electronics provides
excellent passive stability ($<$10 fm/s drift) of the cavity
length.

Optical dipole traps connecting the MOT with the cavity are used
to transport the atoms to the cavity. We employ both travelling
wave and standing wave beam configurations to confine the atoms.
In the former case, the atoms are confined transversely and are
guided under the influence of gravity into the cavity. For the
standing-wave configuration, two counter-propagating beams are
used to form a 1-D lattice, and the atoms are confined vertically
at the anti-nodes of the trap beams.  The vertical motion of the
atoms can then be controlled by varying the difference frequency
$\delta$ of the beams using two phase-locked acousto-optic
modulators, which creates a `walking-wave' with velocity
$v=\lambda \delta/2$, where $\lambda$ is the trapping wavelength
\cite{conveyor}.  The optical traps are loaded directly from the
MOT, following sub-doppler cooling of the atoms to $\sim $4
$\mu$K.

In our first experiment, the trapping laser beam is generated from
a diode laser operating at 782.5 nm (2.25 nm above of the rubidium
D2 line).  A single travelling wave of 16 mW of laser power is
focused to a $1/e^2$ intensity waist of $w_0$ = 30 $\mu$m at the
cavity. This yields a transverse trap depth of 72 $\mu$K at the
cavity. At the location of the MOT, the transverse trap depth is
only 4 $\mu$K due to the divergence of the beam to a waist of 130
$\mu$m. We load up to 10$\times10^6$ $^{87}$Rb atoms into our MOT,
and typically 10$\%$ are transferred into the optical guide. After
the MOT light is extinguished, the atoms fall under gravity
towards the cavity while being compressed transversely in the
guide.

The atoms delivered to the cavity are detected as modifications in
the transmission of a weak cavity probe beam, tuned to the
5S$_{1/2}$F$=2\rightarrow$ 5P$_{3/2}$F'=3 transition in $^{87}$Rb.
For our cavity, the critical photon number (the average number of
photons producing an intracavity intensity of I$_{sat}$) is
$\langle m_0 \rangle=\Gamma^2/(8g_0^2)= 0.006$, and single atoms
are readily detected with probe strengths up to 1.9 pW
corresponding to intracavity fields of $\langle n \rangle\sim1$ .
The probe beam is detected with balanced-heterodyne detection with
typical bandwidth of 30 kHz. The cavity length is adjusted such
that the cavity's resonant frequency $\omega_{c}$ is typically
detuned $\delta_{c} = (2\pi)4\times10^6$ rad/s below both the
atom's ($\omega_{a}$) and the probe beam's ($\omega_{l}$)
frequency.

\begin{figure}
\includegraphics{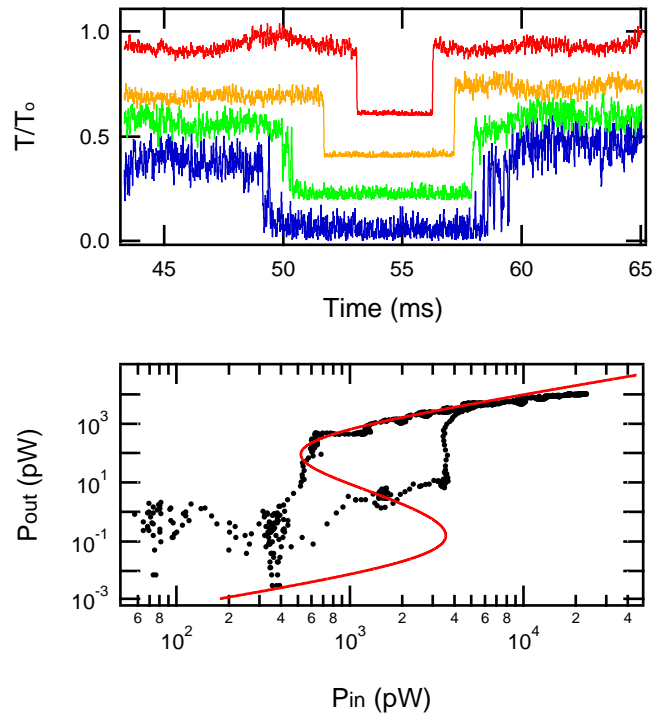}
\caption{(top) The transmission of four different probe beam
powers (2, 6.4, 20, 30 pW from bottom to top) are plotted vs. time
as the atomic cloud is guided though the cavity by the FORT beam.
The graphs are offset by 0.25 each for clarity. (bottom) A plot of
cavity output power vs. input power.  The data was collected in 1
ms while the center of the atomic cloud overlapped with the cavity
mode.  The curve shows a theoretical plot of output vs. input
power given by the optical bistability equation with a
cooperativity $C=200$.}
\end{figure}

The cavity transmission is shown in Fig.\ 2 as the atom cloud
falls through the cavity for several different input powers.
Although the atomic density time-profile through the cavity is
approximately gaussian with a  width ($\sim$ 1.5 ms FWHM)
determined by the cloud temperature, the cavity transmission
switches very abruptly and at different times for different
powers.  For the weakest probe powers, the individual atom
transits are observed as spikes in the transmission at the leading
and trailing edges of the cloud.

The abrupt non-linearity in the cavity transmission is due to the
absorptive optical bistability of the system resulting from the
collective interaction of many radiating atoms with the cavity
field. For many atoms, the transmission of the probe beam is given
by the steady state optical bistability equation
$Y=X[(1+2C\chi)^2+(\delta_{c}/\kappa)^2]$ where
$X=(I_o)/(I_{sat}T)$ and $Y=(I_i)/(I_{sat}T)$ are the output and
input intensities, $I_{i,o}$, normalized to the saturation
intensity $I_{sat}$ and cavity transmission $T$ and $\chi=3
\ln[((1+\sqrt(1+8X/3))/2)]/(2X)$\cite{Orozco}. For $C\geq15$, this
equation predicts two stable output powers for a given input power
as shown in the theoretical trace of Fig.\ 2b as well as
hysteresis in the switching power due to the unstable branch with
negative slope.

To measure the bistability directly, we allow the atomic cloud to
fall into the cavity with the probe beam off, then we quickly ramp
the  probe up to a high value and then down to zero again while
the atoms are in the cavity. The results are shown in Fig.\ 2b.,
where the hysteresis is clearly evident in the 10-fold difference
in the switching power. This data is taken from a single
experimental run in real time. The data agrees very well with the
theoretical curve for high input powers.  The agreement at lower
power is less satisfactory although the noise floor of the
heterodyne detection complicates the analysis of these data.

Coming back to the traces in Fig.\ 2a, the different switching
times can be understood as follows: as the atomic cloud falls
through the cavity, the intracavity atomic density (and hence the
cooperativity $C=NC_1$) first increases then decreases. For
constant input power, the growing cooperativity shifts the
bistability curve to the right until the transmission falls to the
lower branch. Then, as the cloud continues through the cavity, $C$
falls and the transmission then jumps back to the higher branch.
For higher input powers, the switching points occur at higher
intracavity densities and the time window for the drop in
transmission becomes more narrow.

\begin{figure}
\includegraphics{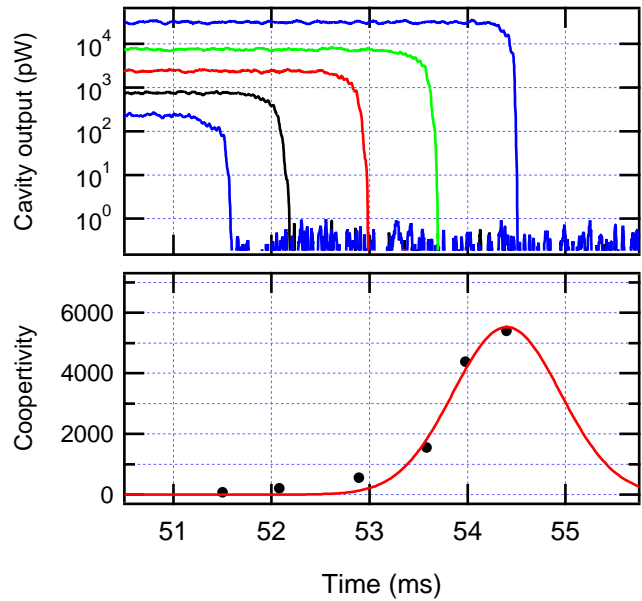}
\caption{(top): Data points from transmission curves of several
different powers are fitted to bistability curves to extract
cooperativity data. From left to right the curves show output
power vs. cooperativity for input powers of 240,758,2400,7580,and
31214 pW. (bottom): The atomic cooperativity vs. time is plotted
as the atomic cloud falls through the cavity.}
\end{figure}

Indeed, the intracavity density at the instant when the cavity
output power drops can be determined by implicitly solving the
bistability equation for $C$ with a known input.  By increasing
the probe power, we can map out the cooperativity and the atom
number vs time. Fig.\ 3 shows data collected in this manner.
Cooperativities of up to 5400 were measured corresponding to an
maximum intracavity atom number of $\sim$100 and an atomic density
$\sim$10$^{10}$ cm$^{-3}$ inside the cavity. For these data, a
Ti:Sapphire laser at 850 nm with 400 mW of power focused to 22
$\mu$m was used to increase the trap depth to 238 $\mu$K, which
increased the number of atoms loaded into the cavity. In addition,
a repumping beam was added perpendicular to the cavity axis to
prevent atoms that are pumped into the $F=1$ ground state by the
cavity probe beam from becoming dark to the probe light
\cite{McKeever}. The maximum switched probe output power is 8 nW,
corresponding to an intracavity intensity of $7\times10^5I_{sat}$.
It is remarkable that such a large intensity can be extinguished
by only 100 atoms.

In our final experiment, we use the Ti:Sapphire laser to generate
two counter-propagating 200 mW beams for the optical lattice. The
beams provide a maximum trap depth of 476 $\mu$K at the cavity,
while at the location of the MOT, the trap depth is only 7 $\mu$K
due to the divergence of the beam waist to 185 $\mu$m.  After the
atoms are loaded into the lattice, we accelerate the occupied
lattice sites down into the cavity mode.  The trapped atoms pass
through the cavity and are brought momentarily to rest. Then they
pass through the cavity again as the lattice velocity is reversed
and they are returned to their original position. The maximum
velocity of the atoms 30 cm/s, and the maximum acceleration
imparted is 1.5$g$. The middle graph in Fig.\ 4 shows the position
and velocity of the lattice sites measured relative to the cavity
axis.

\begin{figure}
\includegraphics{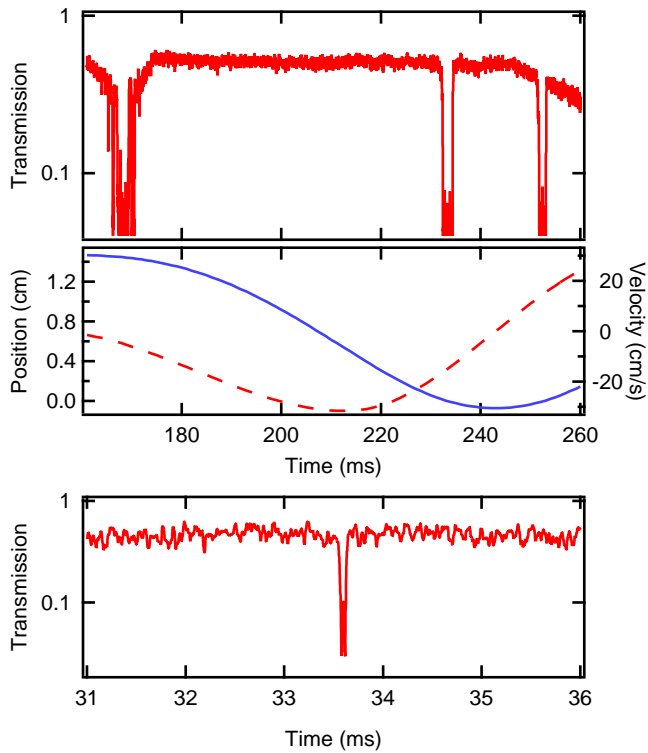}
\caption{(top): Transmission of the cavity probe beam shows a
group of $\sim5$ atoms transported first down and then back up
through the cavity. The first dip in the transmission at 55 ms is
due to unbound atoms as they fall through the cavity due to
gravity. The second and third features are the trapped atoms
moving down and up through the cavity (240 pW probe power).
(middle): Position (solid line) and velocity (dashed line) of the
atoms trapped in the optical lattice. (bottom): Delivery of a
single atom into the cavity mode, arriving 21 ms before the
free-falling atoms released from the MOT (2pW probe power).}
\end{figure}

In Fig.\ 4, the measured transmission through the cavity is shown
for the lattice-transported atoms.  The transmission drops at 120
ms and 140 ms show the atoms on their way down through the cavity
and back up again.  The first feature at 55 ms is due to atoms
that are unbound at lattice sites, but still channelled through
the cavity.  The gradual drop in the baseline transmission for
$t>140 $ is due to the cavity drifting out of resonance due to
heating of the cavity mirrors caused by absorption of the lattice
beams.

With a critical atom number (number of atoms coupled to the cavity
required to alter the atom-cavity response significantly) of
$\langle N_0 \rangle=1/C_1=\Gamma \kappa/(g_0^2)=0.02$, our cavity
is sensitive to single atoms within the mode.  We can load very
few atoms into our MOT by using low light levels and short loading
times. When a single atom is delivered into the cavity mode, the
dressed state of the combined atom-cavity system splits the single
resonant frequency into two peaks, separated by the single photon
Rabi frequency, $2g_0$, which results in a drop in cavity
transmission \cite{jc}.  Fig.\ 4a shows a single atom transported
through the cavity mode. This atom has been accelerated at
$\sim30$ m/s$^2$ and is delivered to the cavity 21 ms before
gravity delivers the unbound atoms.

Currently, our ability to manipulate the atoms in the lattice is
limited by the lifetime of our lattice trap. While the lifetime of
a single travelling wave trap is $\sim2$ s, in the lattice
configuration, the lifetime falls to 104 ms.  We suspect that
power fluctuations in our Ti:Sapphire FORT beams lead to
parametric heating in the lattice configuration \cite{Thomas}.
Indeed, a new solid state pump laser for the Ti:Sapphire has shown
significant lifetime improvements in an independent experiment.

In conclusion, we have realized a cavity QED system with optically
trapped and transported atoms.  Groups of atoms can be
deterministically delivered to the cavity mode. The successful
demonstration of this scalable trapping geometry opens exciting
opportunities in the implementation of quantum gate protocols.
Additionally, the strong coupling parameters of our cavity and the
large maximum intracavity atomic density, provide for a measured
atomic cooperativity, $C$, up to 5400, the largest observed
to-date for atom-cavity systems.

We would like to acknowledge helpful discussions with A. Kuzmich
and L. You. This work was supported by ARDA/NSA/DARPA/ARO (ARO
DAAD19-01-1-0667) and the NSF (PHY-0113831).

\end{document}